\documentstyle[prc,aps,manuscript]{revtex}

\begin{document}

\title{ $(p,\pi^\pm)$ correlations in central
        heavy-ion collisions at $1\div2$ AGeV
        \footnote{Supported by BMBF and GSI Darmstadt} }

\author{ A.B. Larionov \footnote{On leave from RRC "I.V. Kurchatov Institute", 
         123182 Moscow, Russia}, W. Cassing, M. Effenberger, U. Mosel }

\address{ Institut f\"ur Theoretische Physik, Universit\"at Giessen,
          D-35392 Giessen, Germany }

\maketitle

\begin{abstract}
The proton -- charged pion correlated emission is studied in 
the reactions Au (1.06 AGeV) + Au, Ni (1.06 and 1.93 AGeV) + Ni and 
Ni (1.97 AGeV) + Cu within the BUU approach. The associated
invariant mass distributions are shifted to smaller energies with 
respect to the free $\Delta(1232)$ mass distribution due to 
kinematical reasons. We find that the existing and partly conflicting 
experimental data do not allow to draw definite conclusions on the 
in-medium modification of the $\Delta(1232)$.
\end{abstract}

\section{ Introduction }

The properties of the baryon resonances in hot and dense nuclear
matter are the subject of recent theoretical \cite{Eh93,Bass95,KoLi96,%
Teis97,WFN98} and experimental \cite{Hjort97,Eskef98,Pelte99} studies. 
The most important question is, how the centroid energies and the widths 
of the resonances are modified in nuclear matter.

In Ref. \cite{Hjort97} the invariant mass spectra of correlated
$(p,\pi^\pm)$ pairs were measured in collisions of Ni+Cu at 1.97 AGeV.
It is demonstrated in \cite{Hjort97}, that the peaks of the spectra are
shifted to smaller invariant masses by about -50 MeV with respect 
to the free $\Delta$ mass; this shift increases with the collision 
centrality.
Furthermore, in Refs. \cite{Eskef98,Pelte99} the mass distributions of 
the baryon resonances excited in Au+Au and Ni+Ni central collisions 
at energies between 1 and 2 AGeV were deduced experimentally on the basis 
of two methods: (i) by defolding the $p_t$ spectra of charged pions, 
and (ii) from invariant masses of $(p,\pi^\pm)$ pairs. The peaks of 
the $\Delta$-mass distributions extracted in Ref. \cite{Eskef98} are shifted 
on average by $-83$ MeV for Au+Au and $-63$ MeV for Ni+Ni collisions 
with respect to the free $\Delta$-resonance\footnote{ The mass shifts 
reported in Ref. \cite{Eskef98} are given relative to the free $\Delta$ peak 
of 1.210 GeV, which arises from the $\cal B$-function of Ref. \cite{WFN98}
(see also dashed line in Fig. 14).
In our work, the mass distribution of the free $\Delta$ resonance is 
peaked at 1.224 GeV (see solid line line in Fig. 14) and we discuss all 
mass shifts with respect to this value. }.

In this work we have performed BUU transport calculations 
(see Refs. \cite{Teis97,CMMN90} for a general review and Ref. \cite{EBM99} 
for the description of the latest version of the BUU model) 
for collisions of Au+Au at 1.06 AGeV, Ni+Ni at 1.06 and 1.93 AGeV and 
Ni+Cu at 1.97 AGeV in order to study the invariant mass spectrum of 
correlated $(p,\pi^\pm)$ pairs produced by resonance decay.
The aim of this work is twofold:
(i) to study the mechanism of the correlated pair emission, 
and (ii) by comparison with the experimental data try to get  
information on the in-medium modification of the $\Delta$-resonance.

The structure of this paper is as follows: In Sect. II the space-time 
picture of the correlated $(p,\pi^\pm)$ pair emission is studied.
In Sect. III the final correlated $(p,\pi^\pm)$ invariant mass spectra 
are presented and compared to thermodynamic model predictions.
The possibility to extract these spectra within the mixed event technique 
\cite{LHote94} is demonstrated and a comparison to the experimental 
data of Refs. \cite{Hjort97,Eskef98,Pelte99} is performed. 
A summary and conclusions are given in Sect. IV.  

\section{ Space-time picture of the correlated 
          \lowercase{{\large $(p,\pi^\pm)$}} 
          pair emission }

All calculations performed in this work involve the BUU model  
of Ref. \cite{EBM99}, including a momentum-dependent mean field 
with a compressibility modulus $K=280$ MeV as well as vacuum spectral 
functions for all resonances. We shall call this parameterset as
``standard'' below for brevity. Modifications of spectral functions
will be discussed explicitly. Here we only briefly describe the 
most important physical inputs of the BUU model 
(see Ref. \cite{EBM99} for details). 

The mechanism of pion production implemented in the BUU model
includes two steps (besides a small contribution from direct
$NN \rightarrow NN\pi$ processes): In the first step the $\Delta(1232)$ 
and higher baryon resonances are excited in inelastic nucleon-nucleon
collisions: $NN \rightarrow NR$, where $N$ stands for a nucleon 
and $R$ for a resonance. At a collision energy 1$\div$2 AGeV 
dominantly the $NN \rightarrow N\Delta(1232)$ channel determines 
the inelastic part of the total nucleon-nucleon cross section.
The masses of the produced resonances
are chosen according to their vacuum spectral functions 
\begin{equation}
{\cal A}(M) = {2 \over \pi} { M^2 \Gamma(M) \over 
              (M^2-M_{\rm pole}^2)^2 + M^2\Gamma^2(M) }~, 
                                                         \label{FMsq}
\end{equation}
where $M_{\rm pole}$ is the pole mass and $\Gamma(M)$ is the total 
mass-dependent decay width. In the particular case of the 
$\Delta$-resonance $M_{\rm pole}=M_\Delta = 1.232$ GeV, while 
\begin{equation}
\Gamma(M) = \Gamma_\Delta \left({ q \over q_r }\right)^3
{ M_\Delta \over M } { \beta_r^2 + q_r^2 \over \beta_r^2 + q^2 } 
                                                         \label{GammaM}
\end{equation}
is the free space mass-dependent $\Delta$-width with 
$\Gamma_\Delta = 0.118$ GeV, $\beta_r = 0.2$ GeV; $q$ is the pion 
momentum in the rest frame of the $\Delta$ and $q_r$ is the value of 
$q$ at $M = M_\Delta$.

In the second step the resonance decays into a pion and a nucleon:
$R \rightarrow \pi N$. The resonance decay at a given time step is 
simulated by Monte-Carlo using the free space mass-dependent 
decay width $\Gamma(M)$. The Pauli blocking for the outgoing nucleon 
is taken into account in our calculations (c.f. Ref. \cite{Engel94}).
The life time of the $\Delta$-resonance at its pole mass with respect to 
the pion emission is, therefore, $\tau_{\Delta \rightarrow \pi N} = 
1/\Gamma_{\Delta\rightarrow \pi N}\sim 2$ fm/c, where 
$\Gamma_{\Delta\rightarrow \pi N}$ is the in-medium decay width of the 
$\Delta$-resonance, which is less than $\Gamma(M)$ due to Pauli 
blocking \cite{EHTM97}.
Furthermore, the produced pion can be reabsorbed: $\pi N \rightarrow \Delta$. 
Since the average life time of the
$\Delta$ is much less than the characteristic time ($\sim 40$ fm/c)
of a central heavy-ion collision at 1$\div$2 AGeV and the pion 
mean-free-path is quite small ($\lambda_\pi \sim 1$ fm), a chain 
of processes $\Delta_1 \rightarrow \pi_1 N_1$,
$\pi_1 N_2 \rightarrow \Delta_2$, $\Delta_2 \rightarrow \pi_3 N_3$, ...
develops. Moreover, the pion and nucleon rescattering on nucleons
strongly reduces the probability for the decayed $\Delta$ to be ``visible''.
Therefore, only a very small part of $\Delta$-resonances  excited in the
nucleus-nucleus collision, which is not absorbed in the process 
$\Delta N \rightarrow N N$, can be observed by looking at their decay 
products -- i.e. correlated $(\pi,N)$ pairs.
We define a proton-pion pair as being correlated, if this pair
originates from the same resonance and both the proton and the neutron
don't rescatter anymore. In other words, we will always consider
only {\it observable} correlated pairs. 
For the collisions under study we have obtained numerically on average
1$\div$4 $(p,\pi^-)$ and 2$\div$6 $(p,\pi^+)$ correlated pairs per
event in a calculation without acceptance cut (Table 1). 
After filtering through the central drift chamber (CDC) acceptance 
of the FOPI Collaboration (for Au+Au and Ni+Ni reactions) the number 
of correlated pairs decreases by a factor of 3$\div$4. 
The CDC acceptance (see Refs. \cite{Eskef98,Pelte97}) 
was simulated by selecting only particles in the interval 
$32^o< \Theta_{lab} < 150^o$ and, for $\pi^+$, additionally the
cut $p_{lab} < 0.65$ GeV/c was applied.
In Table 1 we present the numbers $N_{(p,\pi^-)}^{corr}$ and
$N_{(p,\pi^+)}^{corr}$ of correlated $(p,\pi^-)$ and $(p,\pi^+)$
pairs and corresponding ratios  
$ r^\pm = N_{(p,\pi^\pm)}^{corr} /  N_{(p,\pi^\pm)}^{ran} $.
Here $N_{(p,\pi^\pm)}^{ran}$ is the number of ``random'' pairs,
i.e. those pairs which are composed of a proton and a pion 
from the same event, but {\it not} emitted from the same resonance:
\begin{equation}
N_{(p,\pi^\pm)}^{ran} = N_p N_{\pi^\pm} 
                      - N_{(p,\pi^\pm)}^{corr}~,      \label{randef}
\end{equation}
where  $N_p$ and $N_{\pi^\pm}$ are the total numbers of emitted 
protons and pions, respectively. 
The ratios $r^\pm$ are quite small ($\sim 10^{-3} \div 10^{-2}$) 
which makes the separation of the true signal from the background 
(see Sect. III B) difficult.

In Fig. 1 we show the time evolution of the nucleon density profile 
and of the radial distribution of the correlated 
$(p,\pi^\pm)$ pairs emitted during the time intervals from $t_i-1$ fm/c 
to $t_i+1$ fm/c, where $t_i$=14, 18, 28 and 38 fm/c, for the Au+Au
system at b=0. Most of the pairs are emitted from the periphery of the 
system, where the density is low (cf. Fig. 3).
We see from Fig. 1, that the intensity of the emission
reaches a maximum at $t \simeq 20$ fm/c and then slowly decreases.
This is consistent with the pion production rate, which reaches
a maximum at t=20 fm/c (cf. Fig. 11 from Ref. \cite{Homb98}).
Thus, pairs are emitted dominantly at $20 \leq t \leq 30$ fm/c.

In Fig. 2 the radial dependence of the temperature is presented
for a central collision of Au+Au at various times accompanied by the 
radial distribution of the correlated $(p,\pi^\pm)$ pairs.
The local temperature was determined on a cubic lattice of 1 fm
grid size introduced in the center-of-mass system of the colliding nuclei. 
A finite temperature Fermi distribution has been adjusted such as to 
obtain the correct baryon density and average baryon kinetic energy in 
the local rest frame of the matter element in the lattice cell. 
During the time period of intensive pair emission, $20 \leq t \leq
30$ fm/c, the temperature at the periphery of the system decreases
from 70 MeV to 30 MeV. This implies, that the description of the freeze-out
of correlated $(p,\pi^\pm)$ pairs by some universal value of the
temperature is quite schematic. The same conclusion can
be obtained with respect to the freeze-out density, which is varying 
strongly during the period of intensive pair emission (Fig. 1).
Nevertheless, one can estimate some average value of the freeze-out
density and temperature for a comparison with macroscopic thermodynamic
models (cf. Refs. \cite{WFN98,Cley99}). 
Fig. 3 shows the distributions of the correlated $(p,\pi)$ 
($\pi=\pi^-,\pi^0,\pi^+$) pairs 
in the freeze-out temperature-density plane for different reactions. 
It is interesting that these distributions are very similar for all
colliding systems regardless of the system size and, moreover,
the collision energy. We always see a doubly peaked structure:
at $T_{f.o.}\approx0$ and $\rho_{f.o.}\approx0$, 
that corresponds to the decay of $\Delta$-resonances in vacuum, and at 
$T_{f.o.}\approx23$ MeV and $\rho_{f.o.}\approx0.14\rho_0$,
where $\rho_0=0.16$ fm$^{-3}$ is the equilibrium nuclear matter density.
In Table 2 the mean values and dispersions of the freeze-out parameters
are given for all reactions. The distributions are quite broad\footnote{
The numerical tables with the distributions of the $(p,\pi)$ pairs
in the freeze-out temperature-density are available from the authors
as data files.} and the mean values of the freeze-out density and 
temperature are essentially higher than the maxima of the actual
distributions in Fig. 3: $\langle \rho_{f.o.} \rangle  \simeq 0.41\rho_0$ 
and $ \langle T_{f.o.} \rangle \simeq 39$ MeV. The baryon chemical 
potential extracted within the thermal model (see next section) 
at $\rho_B = 0.41 \rho_0$ and $T = 39$ MeV is $\mu_B=889$ MeV.
These freeze-out parameters are different from the thermal model 
analysis of Cleymans et al. for Au + Au collisions at 1 AGeV 
(Ref. \cite{Cley99}):  i.e. $T_{f.o.} = 50$ MeV, $\mu_B = 850$ MeV.

\section{ Invariant mass distributions of correlated 
          \lowercase{{\large $(p,\pi^\pm)$}} 
          pairs }

First we discuss the results of the direct analysis of the correlated
$(p,\pi^\pm)$ pairs which can be unambiguously identified in the BUU 
calculation.
In Fig. 4 the invariant mass spectrum of correlated $(p,\pi^\pm)$ pairs
is shown (solid line) for Au+Au central collisions.
This spectrum was 
extracted by stopping the time evolution at $t=40$ fm/c. All the 
correlated pairs present in the system at this time plus the 
pairs produced by the forced decay of resonances were counted.
We have checked that stopping the time evolution at later times
does not influence the extracted spectrum of the pairs within 
the accuracy of our statistics, since for Au + Au central collision 
at 1.06 AGeV only 2\% of all baryon-baryon collisions happen after
40 fm/c. In order to understand, how various time intervals are 
contributing to the total invariant mass spectrum, 
we show also in Fig. 4 the partial spectra given 
by the pairs emitted at $t < 20$ fm/c and at $t < 30$ fm/c. 
We see that, in agreement with the discussion in the previous section, 
the total spectrum is dominantly composed from pairs
emitted in the interval $20 < t < 30$ fm/c and that the low-mass pairs
come primarily from very late times. 

\subsection{ Analysis within the thermodynamical model }

We have analysed the final spectra of correlated $(p,\pi^\pm)$ pairs
using a simple thermodynamical model including nucleons, $\Delta(1232)$
resonances and free pions. For a given temperature $T$ and baryon
density $\rho_B$ the baryon chemical potential $\mu_B$ can be exctracted
from the equation:
\begin{equation}
                 \rho_B = \rho_N + \rho_\Delta~,    \label{rhob}
\end{equation}
where $\rho_N$ and $\rho_\Delta$ are the densities of nucleons 
and deltas:
\begin{eqnarray}
\rho_N &=& 4 \int\,{d^3p \over (2\pi\hbar)^3}\,
{1 \over \exp{(\epsilon_p-\mu_B)/T} + 1}~,          \label{rhon} \\
\rho_\Delta &=& 16 \int\limits_{M_N+M_\pi}^\infty\, dM {\cal A}(M)
\int\,{d^3p \over (2\pi\hbar)^3}\,
{1 \over \exp{(\epsilon_p-\mu_B)/T} + 1}            \label{rhod} 
\end{eqnarray}
with $\epsilon_p = \sqrt{M^2+p^2}$. In Eq. (\ref{rhod}) ${\cal A}(M)$ is the
spectral function of the $\Delta$-resonance given by Eq. (\ref{FMsq}).
The mass distribution of the $\Delta$-resonance at finite chemical
potential $\mu_B$ and temperature $T$ is
\begin{equation}
{ dN_\Delta \over dM } = 
{\cal A}(M) 16 V \int\,{d^3p \over (2\pi\hbar)^3}\,
{1 \over \exp{(\epsilon_p-\mu_B)/T} + 1}~,               \label{dNdeldM}
\end{equation}
where $V$ is the freeze-out volume.

In Fig. 5 we show the invariant mass $(p,\pi)$ 
distributions from the BUU calculations and the $\Delta$-mass
distribution Eq. (\ref{dNdeldM}) at various temperatures 
for the baryon density $\rho_B = \langle \rho_{f.o.} \rangle = 
0.41 \rho_0$. For comparison, the spectral function ${\cal A}(M)$ 
(\ref{FMsq}) is also shown (dot-dashed lines). The high-mass 
tail of the spectral function ($M \geq 1.3$ GeV) is populated very 
weakly. Furthermore, the resulting ($p,\pi$) invariant
mass spectra have a shape quite different from the function ${\cal A}(M)$ 
due to the presence of the integral over the Fermi distribution 
in Eq. (\ref{dNdeldM}).
The peak of the distribution $dN_\Delta / dM$ 
is slightly shifting to higher masses with increasing temperatures. 
We see that the BUU invariant mass spectrum is close to the thermal
one for $T\simeq40\div50$ MeV.  

A better understanding of the BUU results is achieved if we weight
the thermal model calculation with the distribution function 
$\partial^2 N_{pair}(T_{f.o.},\rho_{f.o.})/\partial T_{f.o.}
\partial \rho_{f.o.}$ of the $(p,\pi)$ pairs (Fig. 3).
Then the invariant mass spectrum of the pairs reads as follows
(see Appendix for a derivation):
\begin{equation}
{dN_{pair} \over dM} =
\int\,dT\,\int\,d\rho\,
{ \partial^2 N_{pair}(T,\rho) \over \partial T \partial \rho }\,
{ \Gamma(M) { \partial N_\Delta(M,T,\rho) \over \partial M } \over
  \int\limits_{M_N+M_\pi}^\infty\,dM^\prime\,
\Gamma(M^\prime) { \partial N_\Delta(M^\prime,T,\rho) 
                     \over \partial M^\prime } }~.
                                                   \label{pairaver}
\end{equation}
The distribution (\ref{pairaver}) is shown by solid lines in Fig. 5. 
In  Eq. (\ref{pairaver}) we calculated the integrals for limits
$T=5\div100$ MeV, $\rho=0.1\div2\rho_0$. In this way the pairs
emitted early at nonequilibrium (high density) and late (low density) 
stages of the heavy-ion collision were removed from the analysis. 
The weighted distribution  fits the BUU spectrum better than a calculation 
with fixed temperature and density using Eq. (\ref{dNdeldM}).
Note that Eq. (\ref{pairaver}) has no free parameters:
it only uses as an input the distribution of the emitted pairs
at freeze-out temperature and density. Therefore, the mass distribution of
the emitted pairs from central collisions is consistently described with a 
local thermal equilibrium assumption in agreement with the analysis in 
Refs. \cite{Eskef98,Pelte99}.

\subsection{ Extraction of the correlated ($p,\pi^\pm$) pairs by  
             background subtraction }

The experimental extraction method of the correlated ($p,\pi^\pm$) pairs
is based on background subtraction  from the event-by-event spectrum
\cite{Hjort97,Eskef98,Pelte99,LHote94}. A standard way of the background 
construction is the event mixing technique \cite{LHote94}. We have,
therefore, also prepared mixed BUU events taking protons and pions 
from different events.In the case of Au+Au collisions all events
were taken at b=0 fm. For the reaction Ni+Ni we have taken events 
for $b=1$,2 and 3 fm, and a mixed event can be composed of two
events at different impact parameters\footnote{The background construction 
was done only for the beam energy of 1.93 AGeV in the case of Ni+Ni 
collisions.}. We have checked that an additional restriction of equal 
impact parameters in both events does not influence the result in this case, 
since the collision dynamics changes strongly only for b$\approx5$ fm. 
For the reaction Ni+Cu a mixing of $b=1,2$\dots5 fm events was performed 
selecting event pairs with equal impact parameters only. Then the
difference spectrum agrees with the spectrum of real pairs. However,
for the Ni+Cu reaction, the mixing of the events without imposing any 
condition on the impact parameters produces a difference spectrum
peaked at about 1.15 GeV, i.e. at much smaller invariant mass,
since the background spectrum gets shifted to higher invariant masses.
The conclusion is that the difference spectrum resembles the spectrum
of real pairs only if both events selected for mixing have impact 
parameters within 3 fm. Note, that in our BUU study the reaction plane 
is fixed and, in distinction to the analysis of the experimental data 
in \cite{Eskef98}, we did not rotate our events around the beam axis.  

In Figs. 6-8 the total event-by-event and background spectra (a,c) and
their differences (b,d) are shown for the systems Au+Au, Ni+Ni and Ni+Cu.
For the event mixing procedure we have prepared sets of 
30000, 66000 and 24000 BUU-events for Au+Au, Ni+Ni and Ni+Cu reactions,
respectively. In order to reduce the computational time, these events
were calculated with the Coulomb interaction switched off (for a discussion 
of the Coulomb effects see the next section). The number of mixed events 
is 10 times more for each reaction. The real correlated pairs, selected 
event-by-event, were extracted in parallel from the BUU-events.

The difference spectra (points with errorbars in Figs. 6-8 (b,d)) 
reveal a clear correlation signal at the invariant mass 
$M \simeq 1.2$ GeV, in agreement with the spectra of the correlated 
pairs (solid lines). 

The general agreement between the spectra of real $(p,\pi^\pm)$ pairs
and the difference spectra opens the possibility to identify the spectrum 
of real pairs given by BUU with the experimental difference spectrum.
In this way one can get rid of large statistical errors, which are
mostly due to the background constructed from mixed BUU events.
At the same time, when doing such a comparison, we neglect some
systematical deviations of the difference spectrum, which 
depends on the adopted background construction procedure, from
the spectrum of real correlated pairs, which is unambiguous.
For instance, we see from Figs. 6-8 that, in distinction to the
real spectrum, the difference spectrum can show a negative correlation
and an increased high invariant mass tail. 
Therefore, we will concentrate only on the gross structure 
(like the peak position and the width) of the calculated and measured 
spectra.

\subsection{ Comparison of BUU with the experimental data }

In Figs. 9-11 we show the invariant mass distributions of the correlated
$(p,\pi^-)$ -- (a) and $(p,\pi^+)$ -- (b) pairs for Au+Au at 1.06 AGeV
(Fig. 9), Ni+Ni at 1.06 AGeV (Fig. 10) and Ni+Ni at 1.93 AGeV (Fig. 11)
reactions in comparison to the experimental data of Ref. \cite{Eskef98}. 
The experimental acceptance of the CDC was taken into account in our 
calculations. The theoretical curves are averaged over impact parameter 
in the range $b < 3$ fm, that approximately corresponds to the PM5 
multiplicity bin (see Ref. \cite{Eskef98}). Results with a standard 
parameterset of the BUU model are shown by dotted lines. These results 
are obtained taking into account the Coulomb interactions between
charged particles in the BUU calculation. However, it was supposed 
in the standard calculation, that the momenta of all particles are frozen 
after 40 fm/c and, therefore, the residual Coulomb energy was neglected. 
The effect of the residual Coulomb energy was, furthermore, taken 
into account by rescaling the momenta of particles as:
\begin{equation}
{\bf p} \rightarrow \kappa {\bf p},~~~~~~
\kappa = ( 2 m U_{coul} / p^2 + 1 )^{1/2}~,
\end{equation}
where $U_{coul}$ is the Coulomb energy of a particle at $t=40$ fm/c.
The residual Coulomb energy shifts the spectra of $(p,\pi^-)$ and
$(p,\pi^+)$ pairs to smaller and larger invariant masses, respectively, 
by about 5$\div$10 MeV (dashed lines).

The complete role of the Coulomb interactions can be seen now, e.g. from 
a comparison of the solid line in Fig. 6d with the dashed line in Fig. 9a.
We see that the Coulomb effects are shifting the peak of the $(p,\pi^-)$ 
spectrum in the Au+Au central collisions by about \mbox{-50} MeV. 
The corresponding shift of the $(p,\pi^+)$ spectrum is about +20 MeV 
for the same reaction. At higher beam energies of about 2 AGeV the role 
of the Coulomb interactions becomes negligible for the Ni+Ni and Ni+Cu 
systems.

We have also performed a calculation considering the width of the
$\Delta$-resonance selfconsistently, i.e. taking into account the Pauli
blocking of the decay $\Delta \rightarrow N \pi$, the absorption
$\Delta N \rightarrow N N$ and the rescattering 
$\Delta N \rightarrow \Delta N$. This dynamical width 
of the $\Delta$-resonance then has been used in Eq. (\ref{FMsq}) 
for the spectral function (see Ref. \cite{Eff97} for details).
Moreover, since quite high central densities $\sim 2.5 \rho_0$
are reached in central heavy-ion collisions at $1\div2$ AGeV,
a medium modification of the matrix element for the process 
$\Delta N \rightarrow N N$ simulating the effect of 
a direct three-body absorption (see also Ref. \cite{Engel94}) 
was included as follows:
\begin{equation}
|{\cal M}_{\Delta N \rightarrow N N}|^2 =
\left(1 + 3 {\rho \over \rho_0}\right)
|{\cal M}_{\Delta N \rightarrow N N}^{vac}|^2~.        \label{ddme}
\end{equation}  
The solid lines in Figs. 9-11 show the calculations including the 
selfconsistent $\Delta$-width (wsc) and the density-dependent matrix 
element (ddme) of Eq. (\ref{ddme}).
It gives essentially broader invariant mass spectra and shifts the
peaks to smaller invariant masses, however, still not enough 
to explain the FOPI data on $(p,\pi^+)$ pairs.

One observes, nevertheless, a rather good overall 
agreement of the wsc+ddme calculation to the FOPI data on $(p,\pi^-)$.
For the Ni+Ni collisions at 1.93 AGeV (Fig. 11a), the $(p,\pi^-)$
data reveal a double-humped structure with an additional
peak at low invariant mass, which can be explained by the decay
$\Lambda \rightarrow p \pi^-$ (c.f. Ref. \cite{Hjort97}).
This decay creates a narrow peak at the invariant mass of $1.116$ GeV. 
We obtained on average $\simeq 0.1~ (p,\pi^-)$ pairs per event in central 
collisions of Ni+Ni at 1.93 AGeV and of Ni+Cu at 1.97 AGeV (see below) 
due to the $\Lambda$ decays. The channels $N N \rightarrow \Lambda K N$ and 
$N N \rightarrow \Sigma^0 K N$ followed by $\Sigma^0 \rightarrow \Lambda 
\gamma$ were taken into account in the calculation of the $\Lambda$ 
production. The $\Lambda$ contribution to the $(p,\pi^-)$ spectrum
is shown by the dot-dashed line on top of the wsc+ddme curve
in Fig. 11a. This contribution was parameterized by a gaussian of
width $\sigma=6.4$ MeV/c$^2$ determined by the bin size 20 MeV/c$^2$
of the calculated spectrum. 

Figs. 12,13 show the results for the Ni+Cu reaction at 1.97 AGeV
in comparison to the data from Ref. \cite{Hjort97}. For this reaction
we have performed only the wsc+ddme calculation. The calculated results
are not filtered through the EOS-TPC acceptance. Again, we can observe,
however, a good agreement with the $(p,\pi^-)$ data taking into
account the $\Lambda$ decay. We note, that the relative 
$\Lambda$ contribution is larger in the Ni+Cu reaction than in 
the very similar Ni+Ni reaction. This is caused by the 5$\div$6 times
reduction of the number of $(p,\pi^-)$ pairs due to the $\Lambda$ decays
after filtering through the CDC acceptance, while the number 
of $(p,\pi^\pm)$ pairs produced by the $\Delta$ decays gets reduced
only by a factor of 3$\div$4 (see Table 2). We attribute this different
reduction to a smaller directed $\Lambda$ flow than proton flow (c.f. Ref. 
\cite{LiKo96}). For $(p,\pi^+)$ pairs (Figs. 12b and 13), the calculations
still overpredict the peak position of the $\Delta^{++}$ by about 10 MeV,
but the overall agreement with the EOS-TPC data on $(p,\pi^+)$ pairs is 
better than for the FOPI data.

It was shown in Ref. \cite{Hjort97}, that the peak of the $(p,\pi^+)$
invariant mass distribution shifts to lower invariant masses with 
the centrality of the collision.
Fig. 13 shows, that also in the BUU calculations the same effect is
present since in peripheral nucleus-nucleus collisions 
the emitted proton-pion pairs are mostly due to decays of  
$\Delta$-resonances excited in energetic first-chance nucleon-nucleon 
collisions, where the thermal picture discussed in Sect. III A does 
not apply. We should remark that our statistics is rather poor 
for peripheral collisions. Therefore, the discussed shift in our
calculations in Fig. 13 is more relevant for the average value
of the distribution than for the peak position.

\subsection{ Comparison of the BUU+thermal calculations to 
             the experimental data }

We have studied the predictions of the thermal model using as an input 
the distribution of $(p,\pi)$ pairs at the freeze-out temperature and
density produced by BUU (see Eq. (\ref{pairaver})). This hybrid
approach offers an easy possibility to see the influence 
of the spectral function in the thermal part of the calculation on the 
observed $\Delta$-mass spectrum while {\it retaining in BUU the bare 
spectral function} ${\cal A}$ of Eq. (\ref{FMsq}).  
Besides the calculations with the bare spectral function ${\cal A}$
in the thermal part, we have performed a thermal model analysis 
replacing ${\cal A}$ in Eqs.(\ref{rhod}),(\ref{dNdeldM}) 
by the derivative of the $\pi N$-scattering phase shift
in the $P_{33}$ channel  with respect to the center-of-mass energy of the 
pion and nucleon (Ref. \cite{WFN98}):
\begin{equation}
{\cal B}(E_{cm}) = 2 {\partial \delta_{33}(E_{cm}) 
                       \over
                      \partial E_{cm} }.                \label{B}
\end{equation}
The ${\cal B}$-function (\ref{B}) can be interpreted as a level density 
of the $\pi N$ system (cf. Ref. \cite{Knoll98} and Refs. therein).
The thermal model employing the ${\cal B}$-function (\ref{B}) gives a 
better agreement with the experimental $\pi^0$ spectrum for the Au+Au 
reaction at 1.06 AGeV as shown in Ref. \cite{WFN98}. To clarify the reason 
we compare in Fig. 14 the spectral function ${\cal A}$ (solid line) with 
the weight function ${\cal B}/2\pi$ (dashed line) (see also Fig. 2 in
Ref. \cite{WFN98}). The ${\cal B}$-function yields an increased 
contribution at smaller invariant masses.

Fig. 15 demonstrates the results of the hybrid BUU+thermal calculations
based on Eq. (\ref{pairaver}) for various systems in comparison to the
data from Refs. \cite{Hjort97,Eskef98}. For this comparison, we 
have selected the data on $(p,\pi^+)$, since: (i) the statistics for 
$(p,\pi^+)$ is always better than for $(p,\pi^-)$, and (ii) the $(p,\pi^+)$ 
signal from the $\Delta^{++}$ is not contaminated by the decays of higher 
baryon resonances and by the $\Lambda$ decay. The solid lines in Fig. 15 
show the BUU + thermal calculation with the ${\cal A}$-function. This 
calculation strongly overpredicts the peak positions in the case of Au+Au and 
Ni+Ni reactions (see also dotted lines obtained within the BUU only 
in Figs. 9-11 b). The agreement is somewhat better in the case of 
the Ni+Cu reaction. For the ${\cal B}$-function (dashed lines),
the peak of the theoretical spectrum of the pairs shifts to smaller
invariant masses while the width increases. Thus, the agreement
of the thermal calculation with the FOPI data is improved by the
${\cal B}$-function. The residual discrepancies can be further 
diminished by accounting for the Coulomb effects in the calculations
(c.f. Ref. \cite{Eskef98}). However, the ${\cal B}$-function leads
to a worse agreement for the Ni+Cu reaction, which can be explained 
reasonably well only within the BUU (wsc+ddme) calculation (dotted lines
in Fig. 15). This calculation, however, is still not consistent
with the FOPI data on $(p,\pi^+)$. 
 
\section{ Summary and conclusions }

In this work a study of correlated $(p,\pi^\pm)$ pair emission
from central heavy-ion collisions at energies of 1$\div$2 AGeV 
has been performed within the BUU transport model. 
In agreement with the data \cite{Eskef98}
less than 1$\%$ of the total number of the $(p,\pi^\pm)$ pairs
are correlated. Our calculations give $\approx25\%$ of emitted pions  
in correlations with protons (the rest of pions are produced 
either directly or their correlations are destroyed by proton 
and/or pion rescattering) for Au+Au central collisions at 1.06 AGeV.
This value is lower than the one reported in Ref. \cite{Eskef98}
of $\geq50\%$ of pions correlated with protons. 
Since the total pion multiplicity is also overpredicted by the BUU model 
for the Au+Au reaction by about a factor of 1.7, the number
of correlated pairs per event turns out to be close to the data again.

The correlated pairs originate from $\Delta$-decays in low-density 
regions during an intermediate stage ($t = 20\div30$ fm/c) of the collision. 
The calculated invariant
mass spectra of these pairs have a thermal shape; however, the
real (kinetic) temperature in the system changes quite strongly during
the period of emission. This result is related to a weak sensitivity 
of the shape of the invariant mass spectra to the temperature
in the region $T = 40\div55$ MeV, which follows from a simple 
thermodynamical calculation (see text and Fig. 5). We would like
to stress, that {\it the freeze-out of the pairs happens during an
extended period of time, comparable with the time scale of
the heavy-ion collision itself}. Therefore, the thermodynamical state 
of the system is changing quite strongly during this period. 

The BUU calculation with the selfconsistent treatment of the 
$\Delta$-width and the medium-modified matrix element for the process 
$N \Delta \rightarrow N N$ (see Sect. III C) gives a good overall
agreement with both the FOPI and EOS-TPC data on the $(p,\pi^-)$ 
invariant mass distributions. The calculated $(p,\pi^+)$ spectra
are shifted to higher invariant masses with respect to the
FOPI data, but are in reasonable agreement with the EOS-TPC data.
It is, therefore, an open question if some additional effects 
influencing the $\Delta$ propagation and decay in hot and dense nuclear 
matter are needed to account for the remaining shift in future dynamical 
calculations. In particular, the approach taking into account 
the vacuum $\pi N$ scattering phase
shift (Refs. \cite{WFN98,Dan96}) gives a better agreement 
with the FOPI data, as demonstrated within the thermodynamical model
in Ref. \cite{Eskef98} and in Sect. III D of the present work; however,
it gives a worse description of the EOS data from the BEVALAC.
This conflicting situation calls for new experiments 
on $(p,\pi)$ correlations.

\section*{ Acknowledgments }

We are grateful to E.L. Bratkovskaya, C. Greiner and A.A. Sibirtsev 
for helpful discussions and their interest in this work. Furthermore,
the authors would like to thank D. Pelte for a careful reading of 
the manuscript and helpful advice.

\section*{ Appendix }

In this Appendix we derive Eq. (\ref{pairaver}) for the mass distribution 
of emitted pairs. 

The density of emitted pairs in the
space invariant mass -- freeze-out temperature -- freeze-out density
is (for brevity we drop lower indices at $T_{f.o.}$ and $\rho_{f.o.}$):
\begin{equation}
{ \partial^3 N_{pair}(M,T,\rho) \over 
  \partial M \partial T \partial \rho } =
W \Gamma(M) \int\limits_0^\infty dt\,   
{ \partial^3 N_{\Delta}(M,T,\rho,t) \over 
  \partial M \partial T \partial \rho }~,            \label{A1}
\end{equation}
where $\partial^3 N_{\Delta}(M,T,\rho,t) \over 
\partial M \partial T \partial \rho$ is the density of $\Delta$-resonances
in the same space as a function of time, $\Gamma(M)$ is the decay
width of Eq. (\ref{GammaM}), and $W$ is the surviving probability
of an emitted pair assumed to be independent on invariant mass, temperature
and density.
We have:
\begin{equation}
{ \partial^2 N_{pair}(T,\rho) \over \partial T \partial \rho } =
\int\limits_{M_N+M_\pi}^\infty dM\, W \Gamma(M)
\int\limits_0^\infty dt\,
{ \partial^3 N_{\Delta}(M,T,\rho,t) \over 
  \partial M \partial T \partial \rho }~.            \label{A2}
\end{equation}
Assuming local thermal equilibrium we can write:
\begin{equation}
{ \partial^3 N_{\Delta}(M,T,\rho,t) \over 
  \partial M \partial T \partial \rho }   =
{ \partial^2 N_{\Delta}(T,\rho,t) \over \partial T \partial \rho }
{ \partial \tilde N_\Delta(M,T,\rho) \over \partial M }~,   
                                                         \label{A3}
\end{equation}
where
\begin{equation}
{ \partial \tilde N_\Delta(M,T,\rho) \over \partial M } =
{ \partial N_\Delta(M,T,\rho) \over \partial M } 
\left[ 
\int\limits_{M_N+M_\pi}^\infty dM^\prime\, 
{ \partial N_\Delta(M^\prime,T,\rho) \over \partial M^\prime }
\right]^{-1}                                             \label{A4}
\end{equation}
is the mass distribution of $\Delta$-resonances (see Eq. (\ref{dNdeldM}))
normalized to 1, which depends only on the local temperature and density.

Substituting Eqs. (\ref{A3}),(\ref{A4}) into Eqs. (\ref{A1}) and (\ref{A2}),
we obtain the expressions:
\begin{eqnarray}
{ \partial^3 N_{pair}(M,T,\rho) \over 
  \partial M \partial T \partial \rho } & = &
\xi_\Delta(\rho,T)
W \Gamma(M){ \partial \tilde N_\Delta(M,T,\rho) \over \partial M }~,
                                                      \label{A5}   \\
{ \partial^2 N_{pair}(T,\rho) \over \partial T \partial \rho } & = &
\xi_\Delta(\rho,T)
\int\limits_{M_N+M_\pi}^\infty dM\, 
W \Gamma(M){ \partial \tilde N_\Delta(M,T,\rho) \over \partial M }
                                                      \label{A6}
\end{eqnarray}
with
\begin{equation}
\xi_\Delta(\rho,T) =
\int\limits_0^\infty dt\,
{ \partial^2 N_{\Delta}(T,\rho,t) \over \partial T \partial \rho }~.
                                                     \label{A7}
\end{equation}
Using Eqs. (\ref{A5}),(\ref{A6}) and the relation
\begin{equation}
{ d N_{pair} \over d M } = \int dT \int d\rho\,
{ \partial^3 N_{pair}(M,T,\rho) \over 
  \partial M \partial T \partial \rho }               \label{A8}
\end{equation}
it is straightforward to obtain Eq. (\ref{pairaver}).

\newpage

\begin{description}
\item[Table 1] Numbers of correlated $(p,\pi^\pm)$ pairs 
$N_{(p,\pi^\pm)}^{corr}$   
and ratios of numbers of correlated and random 
$(p,\pi^\pm)$ pairs $r^{(\pm)}$ (see text for definitions) for 
central collisions. In the case of Au+Au and Ni+Ni reactions,
the two numbers are given for $N_{(p,\pi^\pm)}^{corr}$
separated by a comma: before and after filtering through the
CDC acceptance. The calculated values of $r^{(\pm)}$ are practically
insensitive to the filtering.
The experimental values of $r^{(\pm)}$
from Ref. \cite{Eskef98} are given in brackets. For Ni+Cu collisions
all results are unfiltered.  
\end{description}

\vspace{0.5cm}

\begin{center}
\begin{tabular}{|c|c|c|c|c|c|}
\hline
System &  Energy &  $N_{(p,\pi^-)}^{corr}$  &  r$^{(-)}$  &  
$N_{(p,\pi^+)}^{corr}$  &  r$^{(+)}$  \\
       &  (AGeV) &   &   (\%)   &     &  (\%)       \\
\hline
Au+Au & 1.06  & 4.0, 1.2 & 0.1 (0.6$\pm$0.2) &  5.7, 2.1 & 
0.3 (0.75$\pm$0.25) \\
Ni+Ni & 1.06  &  1.2, 0.4 & 0.4 (0.75$\pm$0.25) &  2.4, 0.7  &  
0.9 (1.0$\pm$0.3) \\
Ni+Ni & 1.93  &  2.5$^*$, 0.6  &  0.4 (0.6$\pm$0.2)  &  4.9, 1.2  &  
0.9 (1.05$\pm$0.3) \\
Ni+Cu & 1.97  &  2.7$^*$  &  0.3            &  4.9  &
0.7  \\
\hline   
\end{tabular}
\end{center}

\vspace{0.5cm} $^*$ including 0.1 $(p,\pi^-)$ pair due to $\Lambda$ decays

\vspace{2cm}

\begin{description}
\item[Table 2] Mean values and dispersions of the freeze-out 
density and temperature for central collisions.
\end{description}

\vspace{0.5cm}

\begin{center}
\begin{tabular}{|c|c|c|c|c|c|}
\hline
System & Energy &   $\langle\rho_{f.o.}\rangle$  &  $\sigma_\rho$  &
$\langle T_{f.o.}\rangle$  &  $\sigma_T$  \\
       & (AGeV) &   ($\rho_0$)                   &  ($\rho_0$)     &
(MeV)                      &  (MeV)       \\
\hline
Au+Au  & 1.06   &   0.41   &  0.39  &  39.0  &  21.0  \\
Ni+Ni  & 1.06   &   0.45   &  0.42  &  39.4  &  25.6  \\
Ni+Ni  & 1.93   &   0.40   &  0.42  &  37.8  &  24.9  \\
Ni+Cu  & 1.97   &   0.41   &  0.42  &  37.9  &  24.9  \\
\hline   
\end{tabular}
\end{center}

\newpage

\section*{ Figure captions }

\begin{description}

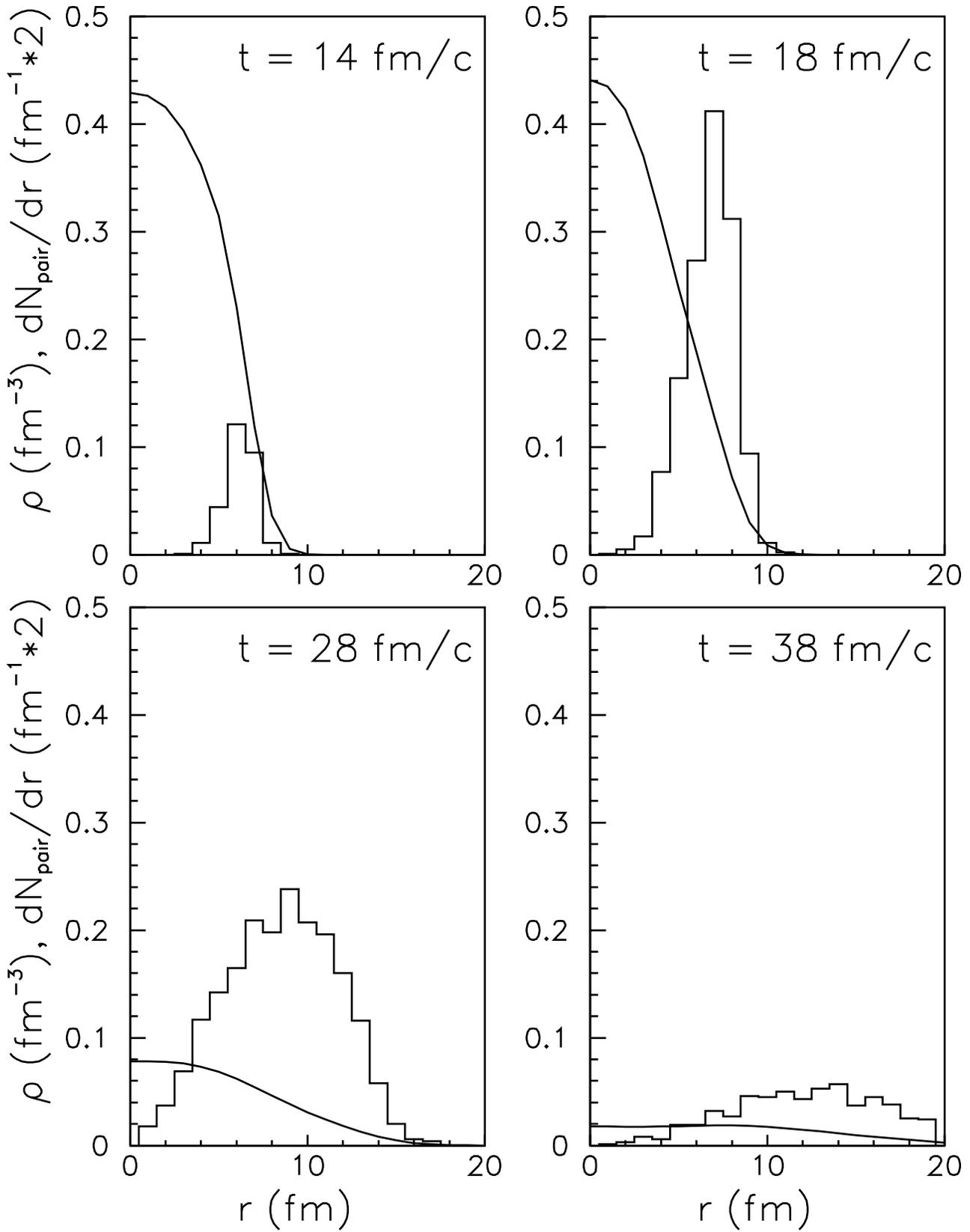
\item[Fig. 1] Radial distributions of the correlated $(p,\pi^\pm)$
pairs emitted during the time interval $\Delta t = 2$ fm/c (histograms)
and the baryon density profiles (solid lines) at various times
for a central collision of Au+Au at 1.06 AGeV. Pairs are selected
without acceptance cuts.

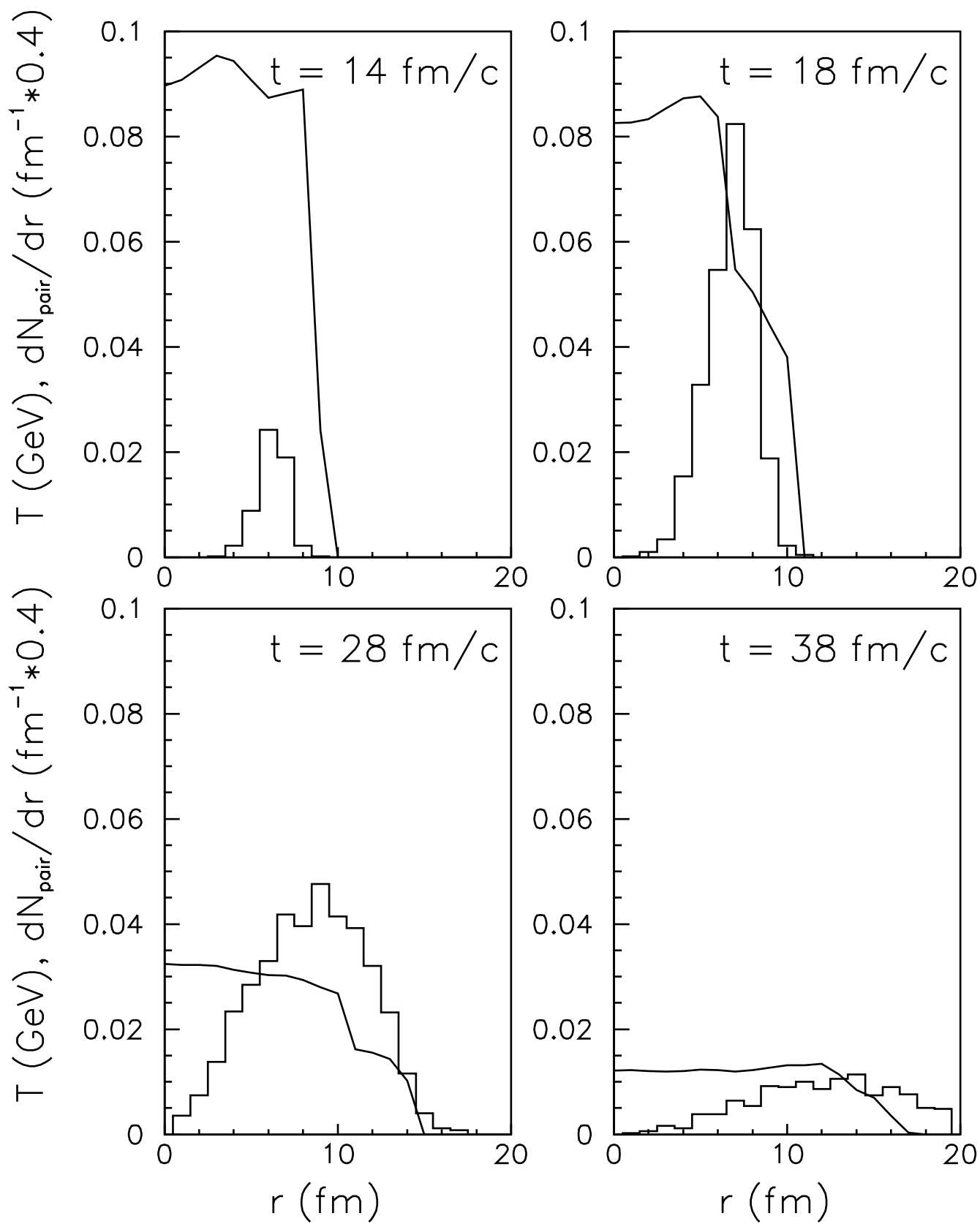
\item[Fig. 2] The same as Fig. 1, but with the kinetic temperature 
shown by solid lines.

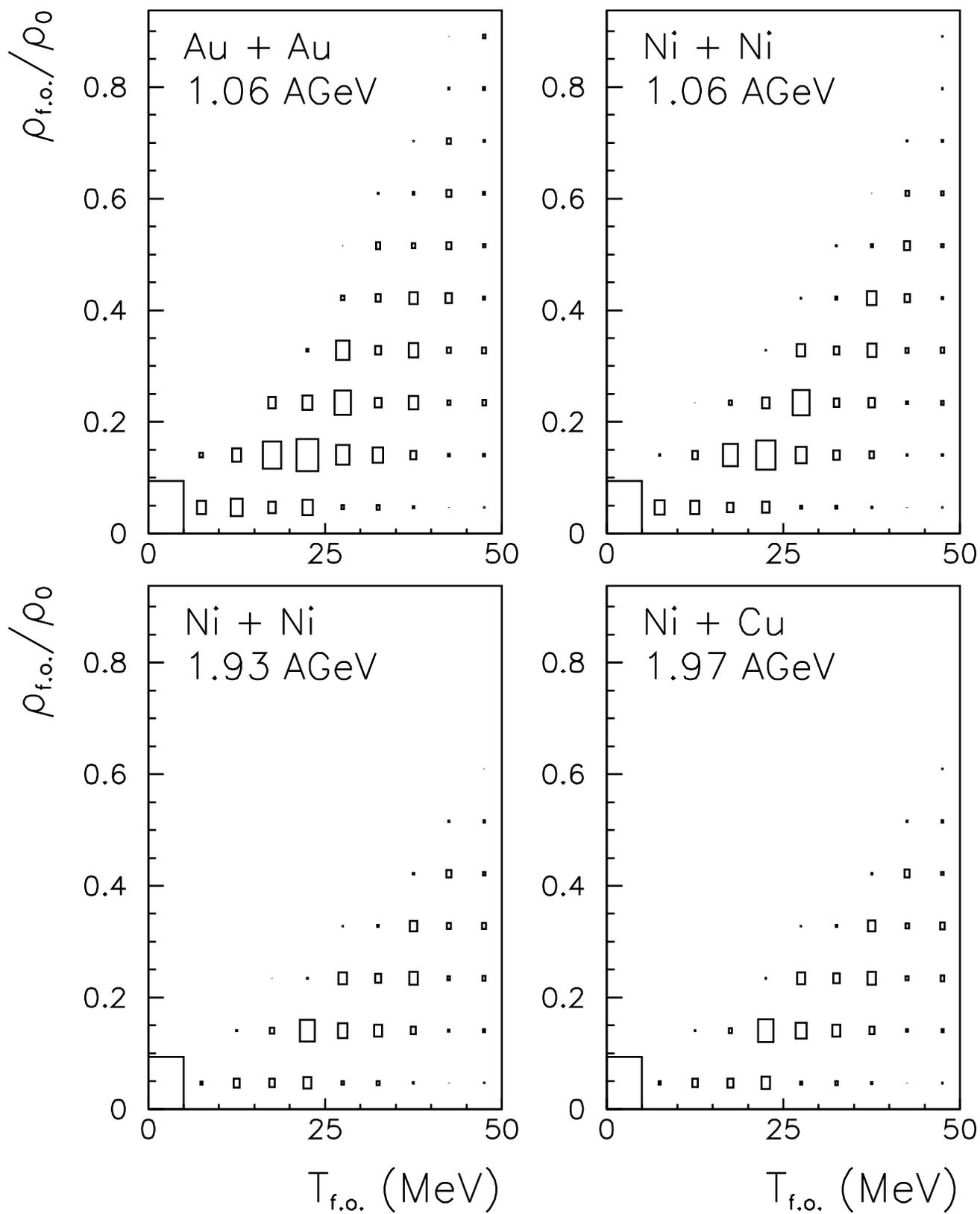
\item[Fig. 3] Freeze-out temperature-density distribution 
$\partial^2 N_{pair}(T_{f.o.},\rho_{f.o.})/\partial T_{f.o.}
\partial \rho_{f.o.}$ of the correlated $(p,\pi)$ pairs 
for central collisions of various systems. The distribution (in a.u.) at 
a given point ($T_{f.o.},\rho_{f.o.}$) is proportional to the size of 
the box. No acceptance cuts were applied.

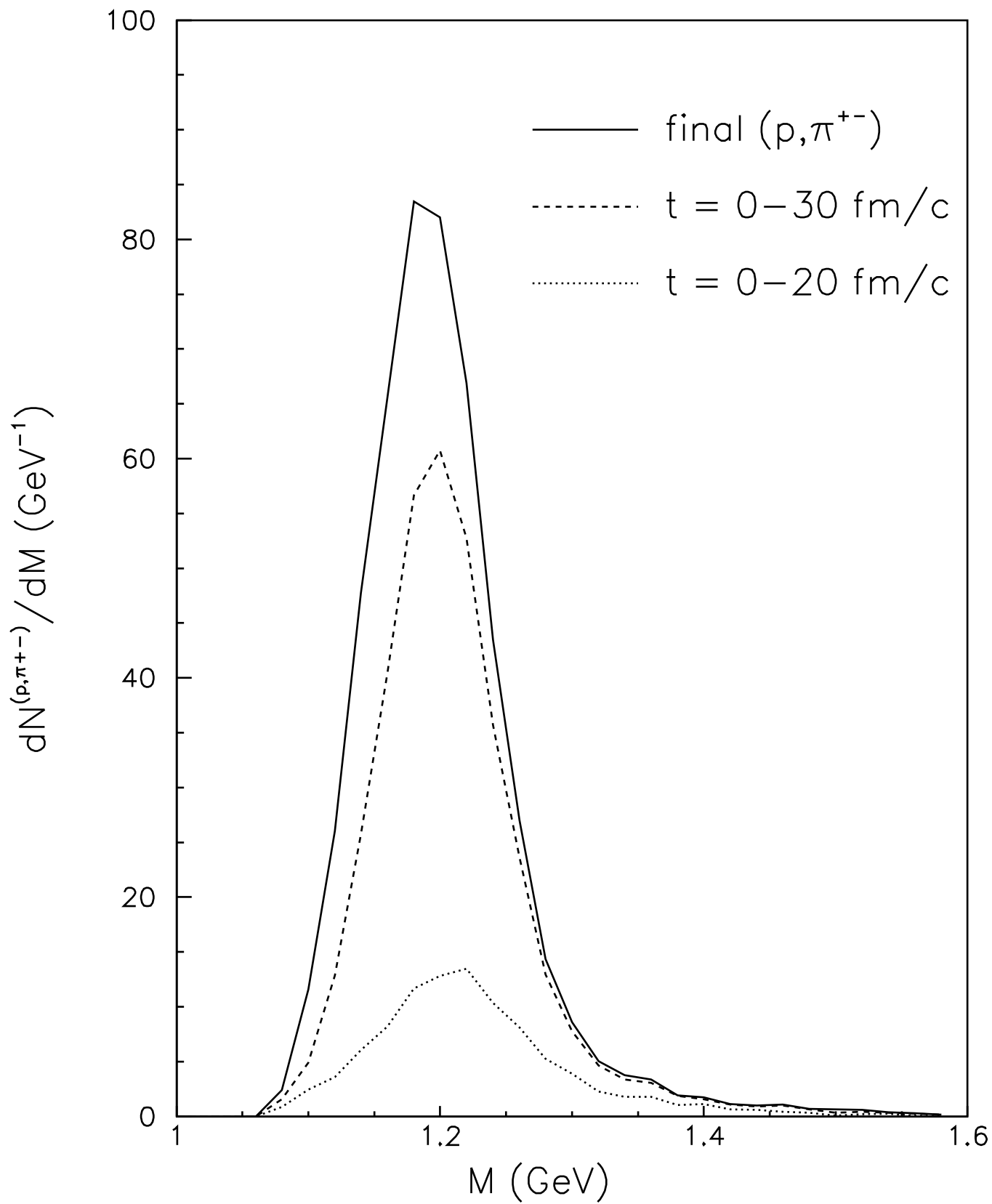
\item[Fig. 4] Invariant mass distribution of the correlated $(p,\pi^\pm)$
pairs emitted at the time intervals $t = 0\div20$ fm/c (dotted line),
$t = 0\div30$ fm/c (dashed line) and the total spectrum after induced
decay of residual $\Delta$'s at $t = 40$ fm/c (solid line) for a central
collision of Au+Au at 1.06 AGeV. No acceptance cuts were applied.

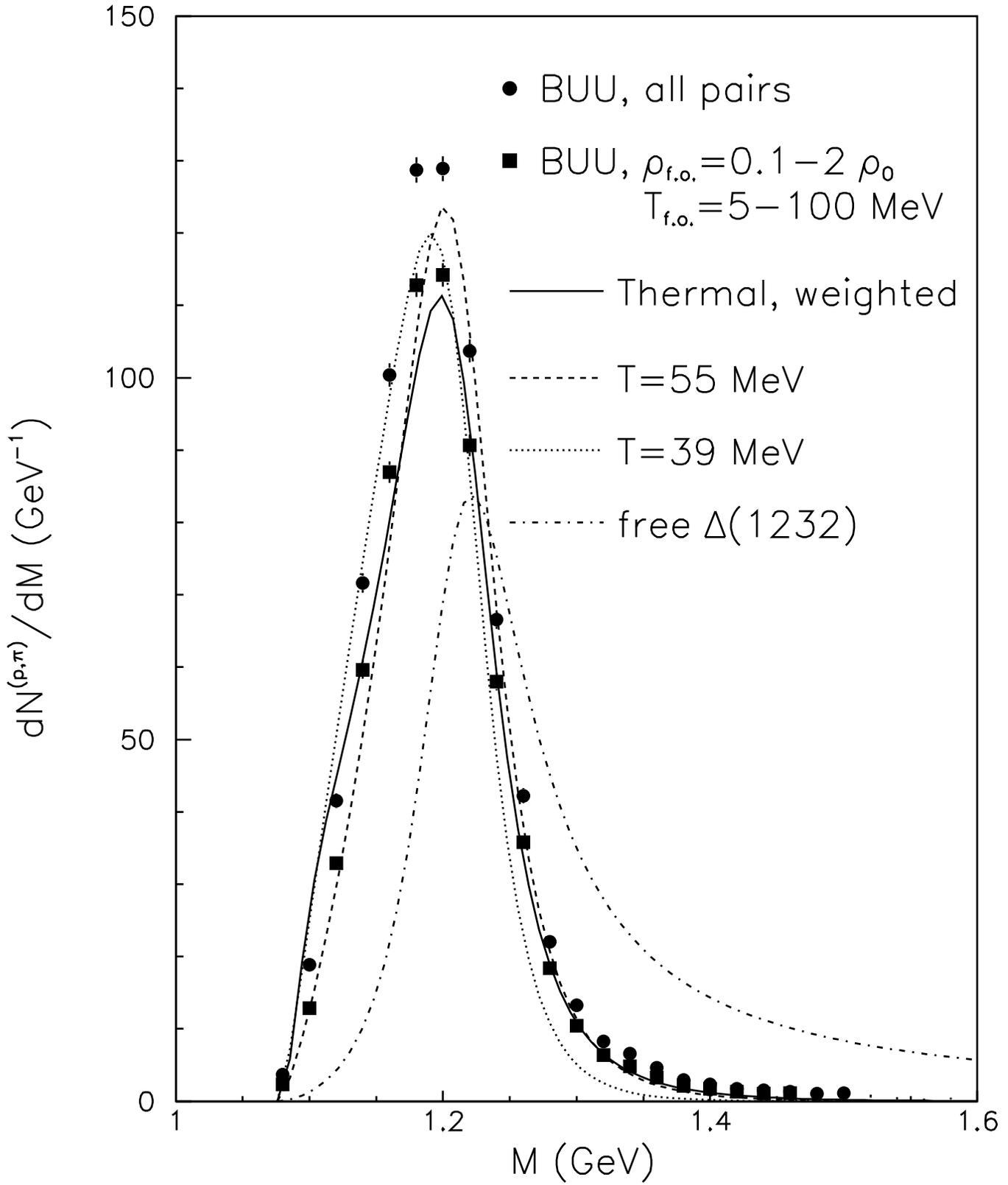
\item[Fig. 5] $(p,\pi)$ ($\pi=\pi^-,\pi^0,\pi^+$) invariant mass
spectra from the BUU calculations for central Au+Au collisions 
at 1.06 AGeV for all pairs (filled circles) and for
pairs emitted at   
$T_{f.o.}=5\div100$ MeV and $\rho_{f.o.}=0.1\div2\rho_0$ 
(filled squares) in comparison to the thermal model calculation
at the temperature $T=39$ MeV and the baryon density 
$\rho_B = 0.41 \rho_0$ (dotted line), $T=55$ MeV and
$\rho_B = 0.41 \rho_0$ (dashed line), 
and weighted with the distribution function 
$\partial^2 N_{pair}(T_{f.o.},\rho_{f.o.})/\partial T_{f.o.}
\partial \rho_{f.o.}$ according to Eq. (\ref{pairaver})
(solid lines). 
The free spectral function ${\cal A}(M)$ 
(Eq. (\ref{FMsq})) is shown by the dash-dotted line.
Curves for fixed temperatures and the free spectral function are
normalized to the number of pairs emitted at 
$T_{f.o.}=5\div100$ MeV and $\rho_{f.o.}=0.1\div2\rho_0$.
No acceptance cuts were applied in selecting the pairs.

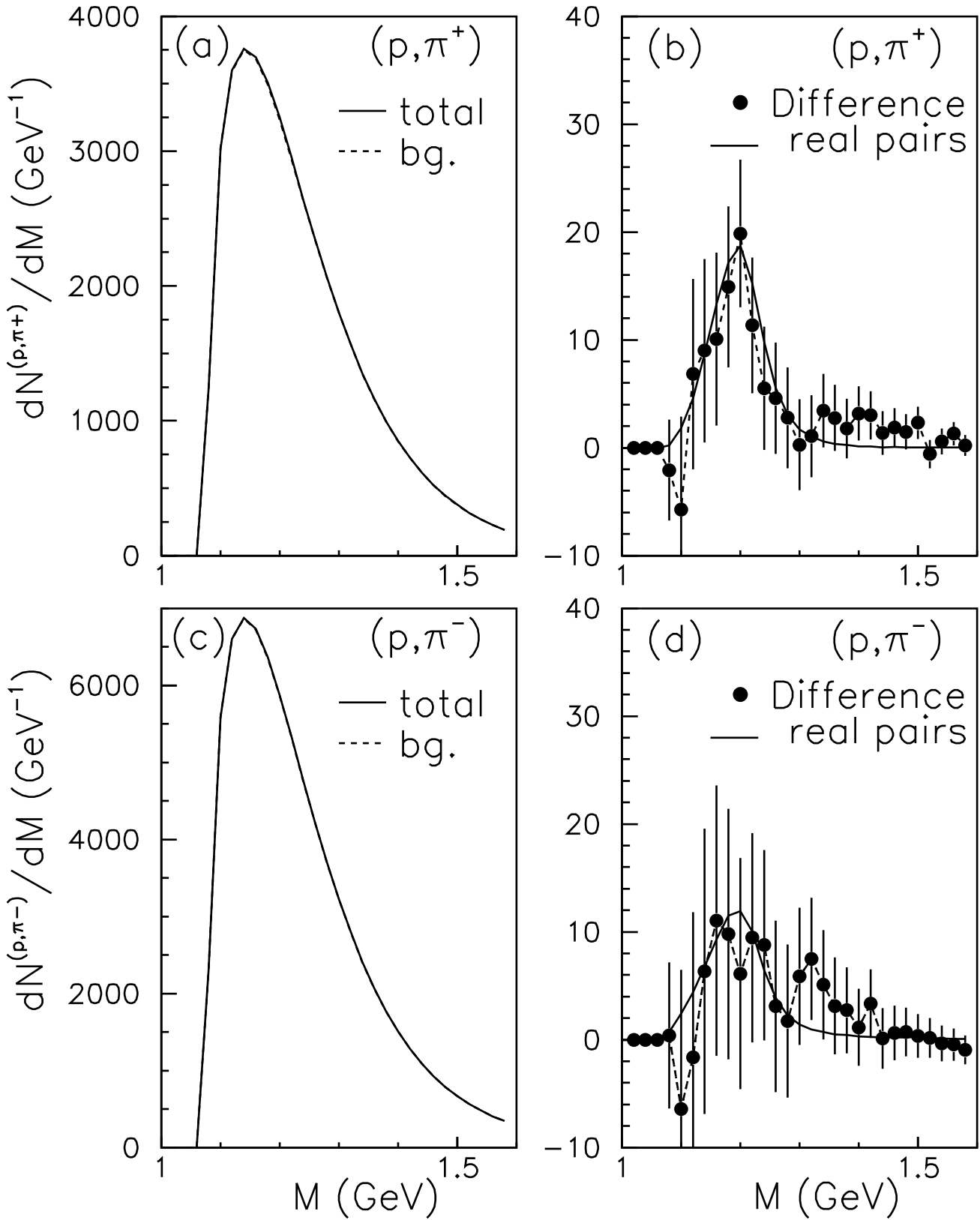
\item[Fig. 6] The left panels show the spectra of all $(p,\pi^+)$ -- (a) and
$(p,\pi^-)$ -- (c) pairs extracted
event-by-event (solid line) and by event mixing (dashed line) from
the BUU calculation for Au+Au at 1.06 AGeV.
The right panels present the corresponding differences between event-by-event 
and mixed-event spectra of $(p,\pi^+)$ -- (b) and $(p,\pi^-)$ -- (d) pairs
(points with errorbars connected by dashed line) and the spectra
of real pairs from the same BUU calculation (solid line).
The particle selection was performed including the CDC acceptance. 

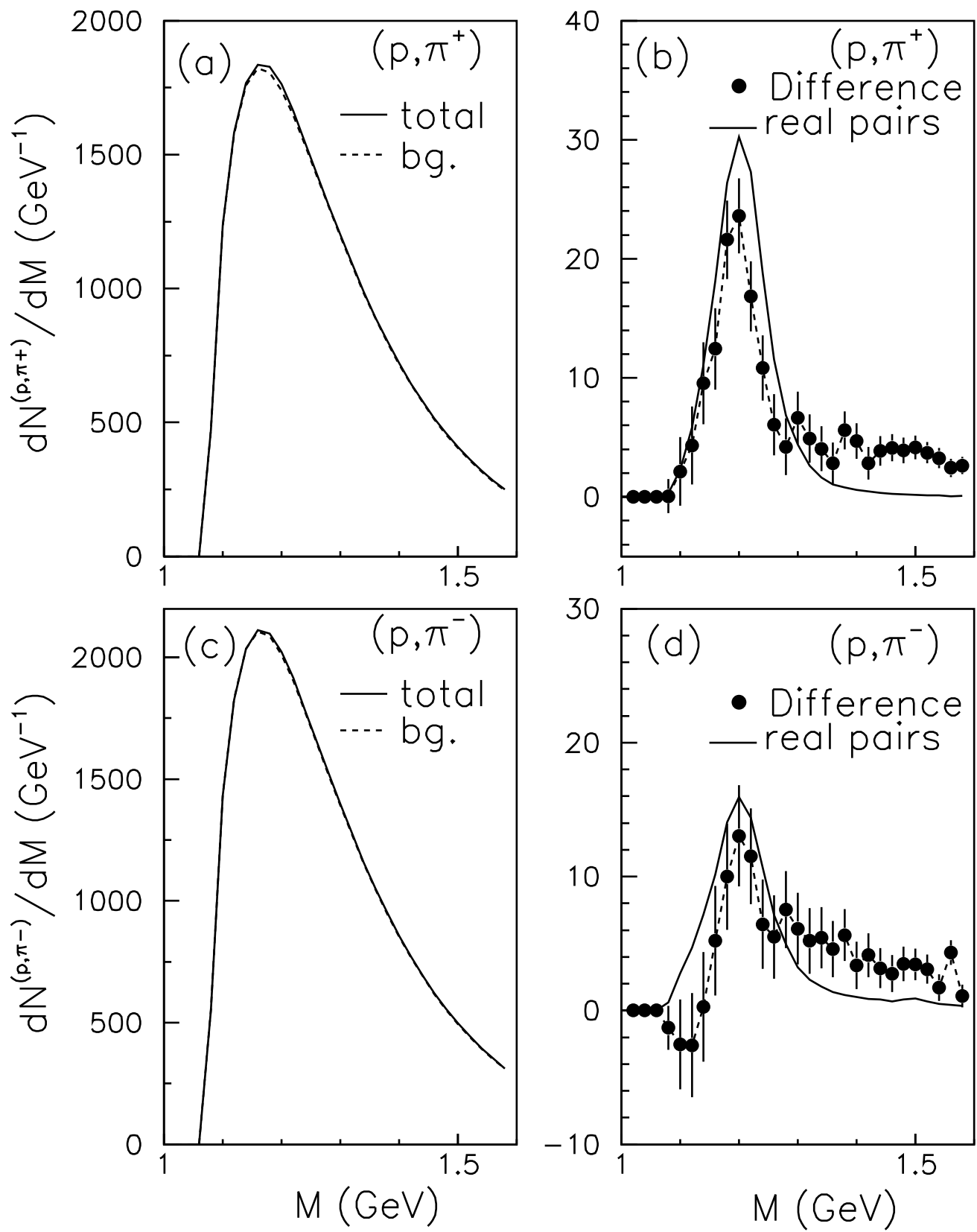
\item[Fig. 7] The same as Fig. 6, but for Ni+Ni collisions at
1.93 AGeV.

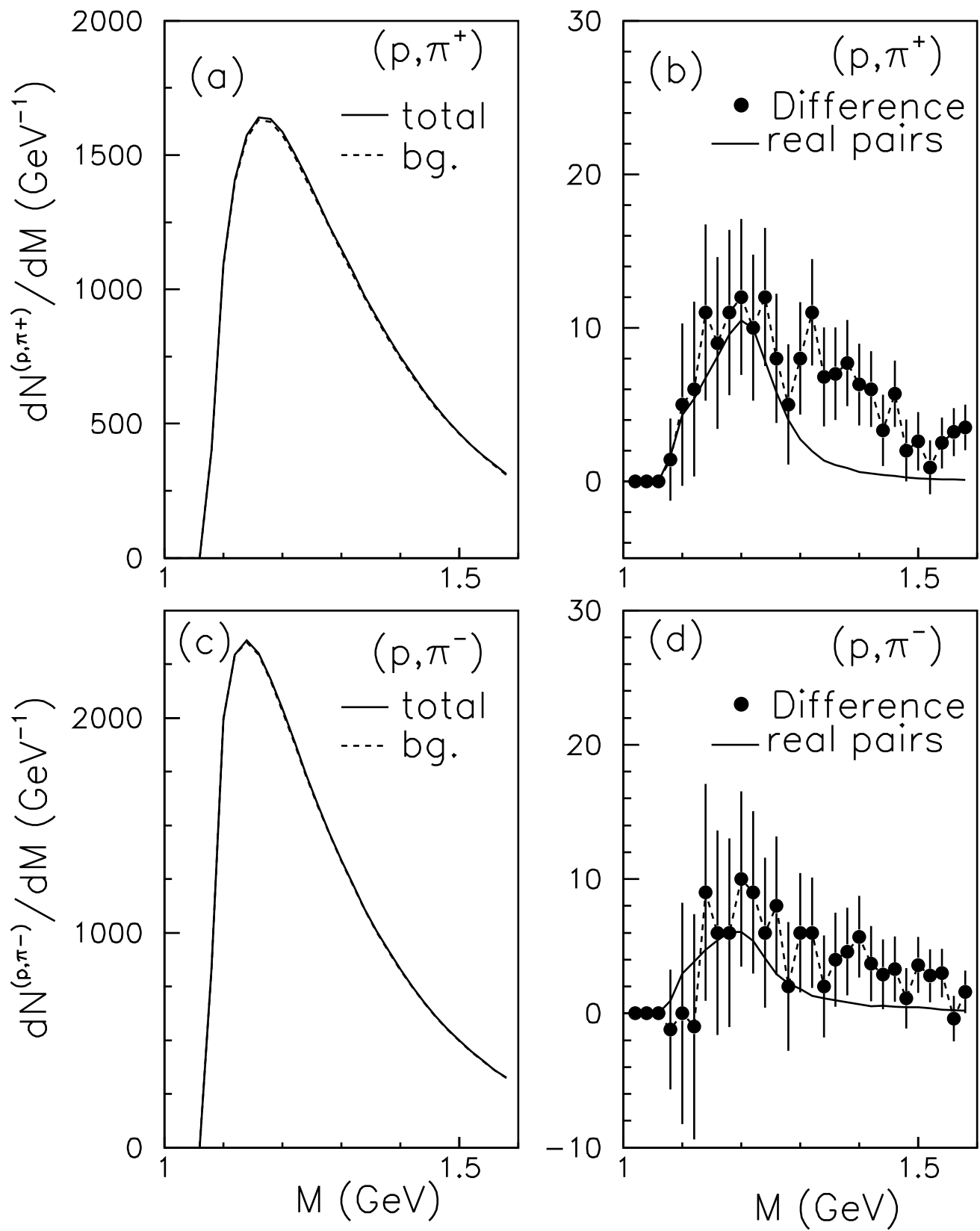
\item[Fig. 8] The same as Fig. 6, but for Ni+Cu collisions
at 1.97 AGeV. Particles are selected in full $4\pi$ acceptance.
The $\Lambda$-decay contribution to the $(p,\pi^-)$ spectra is dropped.

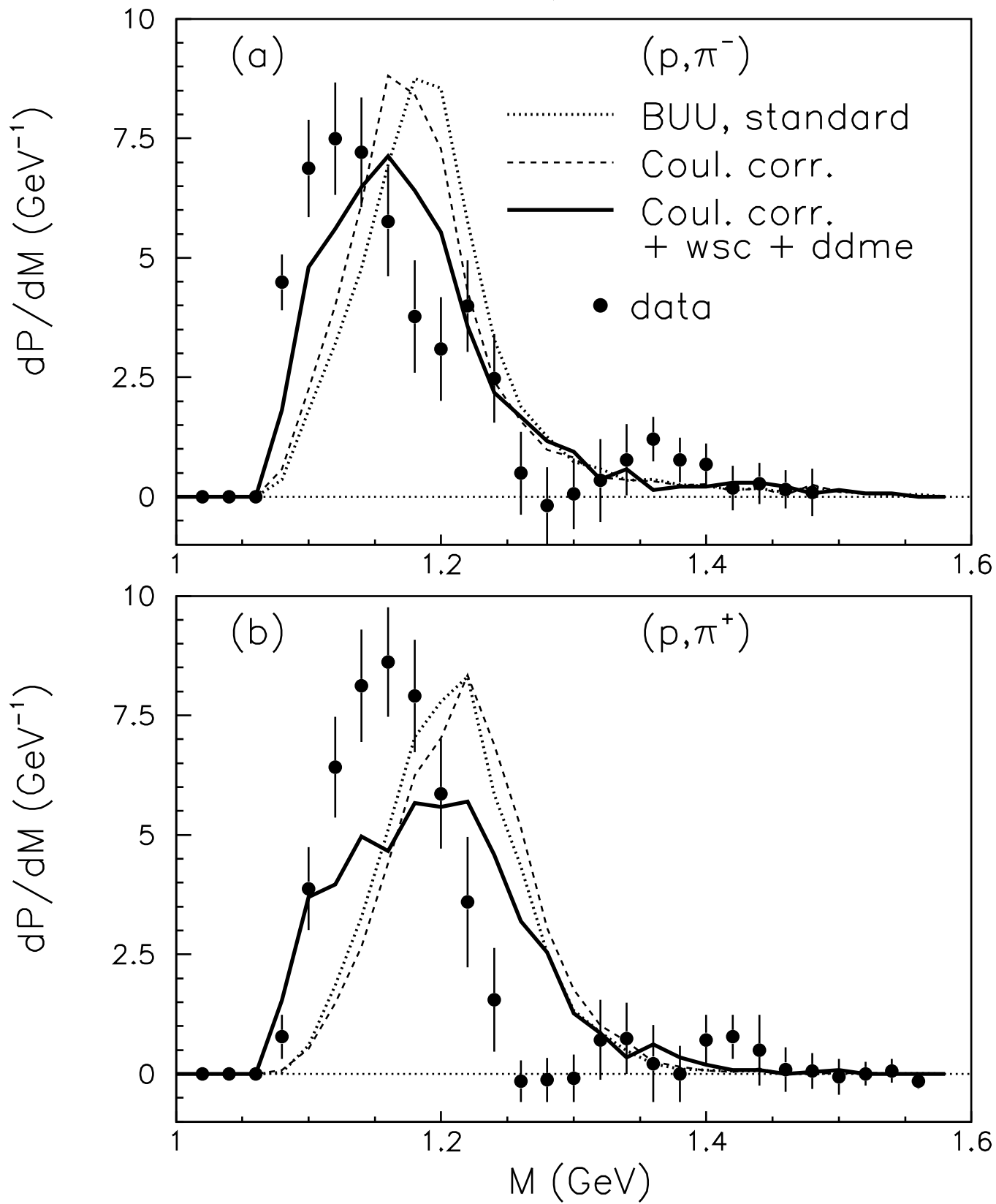
\item[Fig. 9] Invariant mass spectra of $(p,\pi^-)$ (a)
and $(p,\pi^+)$ (b) pairs from central Au+Au collisions at 
1.06 AGeV in comparison to the data from Ref. \cite{Eskef98}.
Dotted lines -- standard BUU parameterset,
dashed lines -- standard BUU with Coulomb final state interaction,
solid lines -- BUU calculation with selfconsistent width of the 
$\Delta$-resonance and a density-dependent matrix element (see text for 
details) plus Coulomb final state interaction. All spectra are 
normalized to unity.

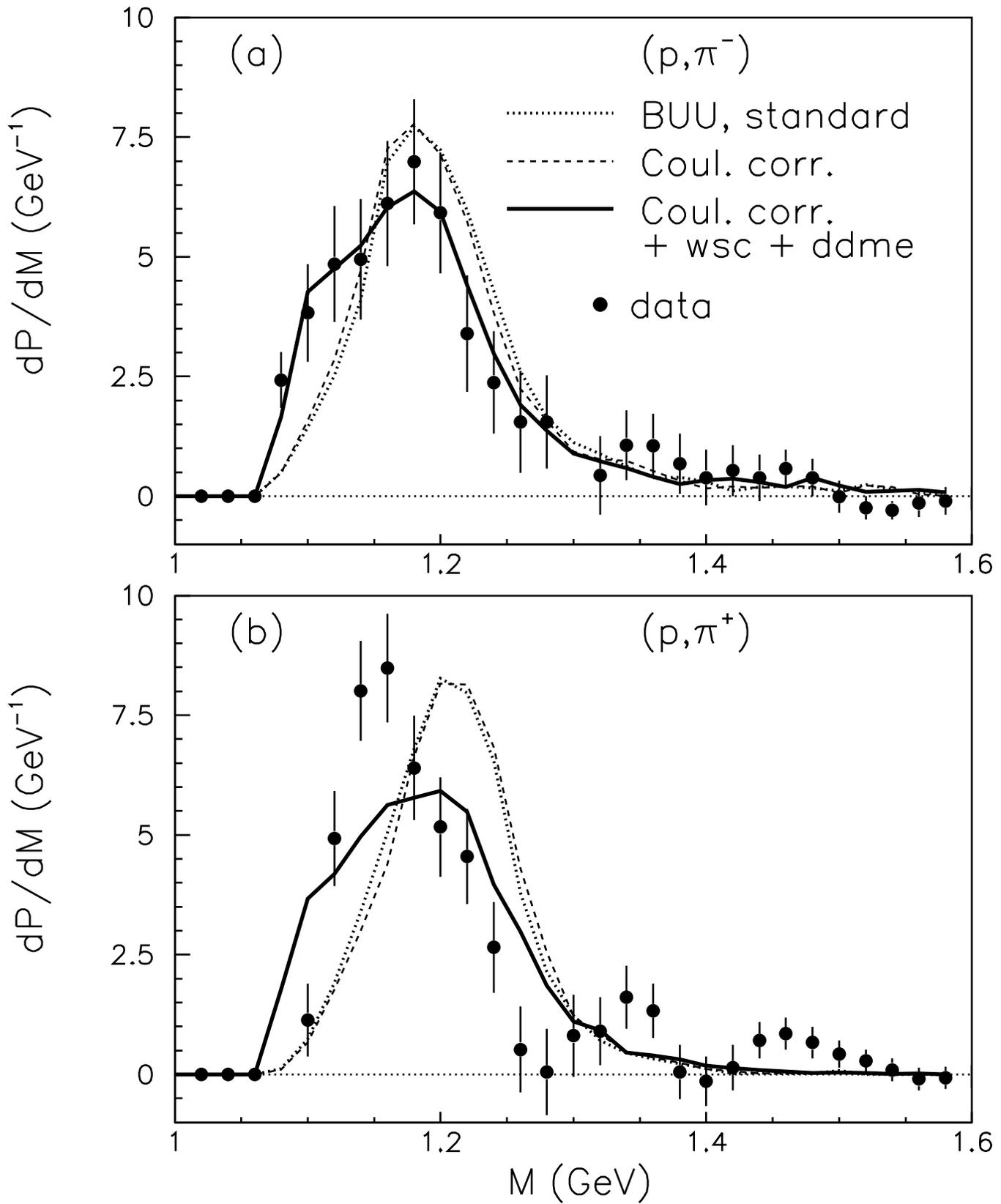
\item[Fig. 10] The same as Fig. 9, but for central collisions of  
Ni+Ni at 1.06 AGeV.

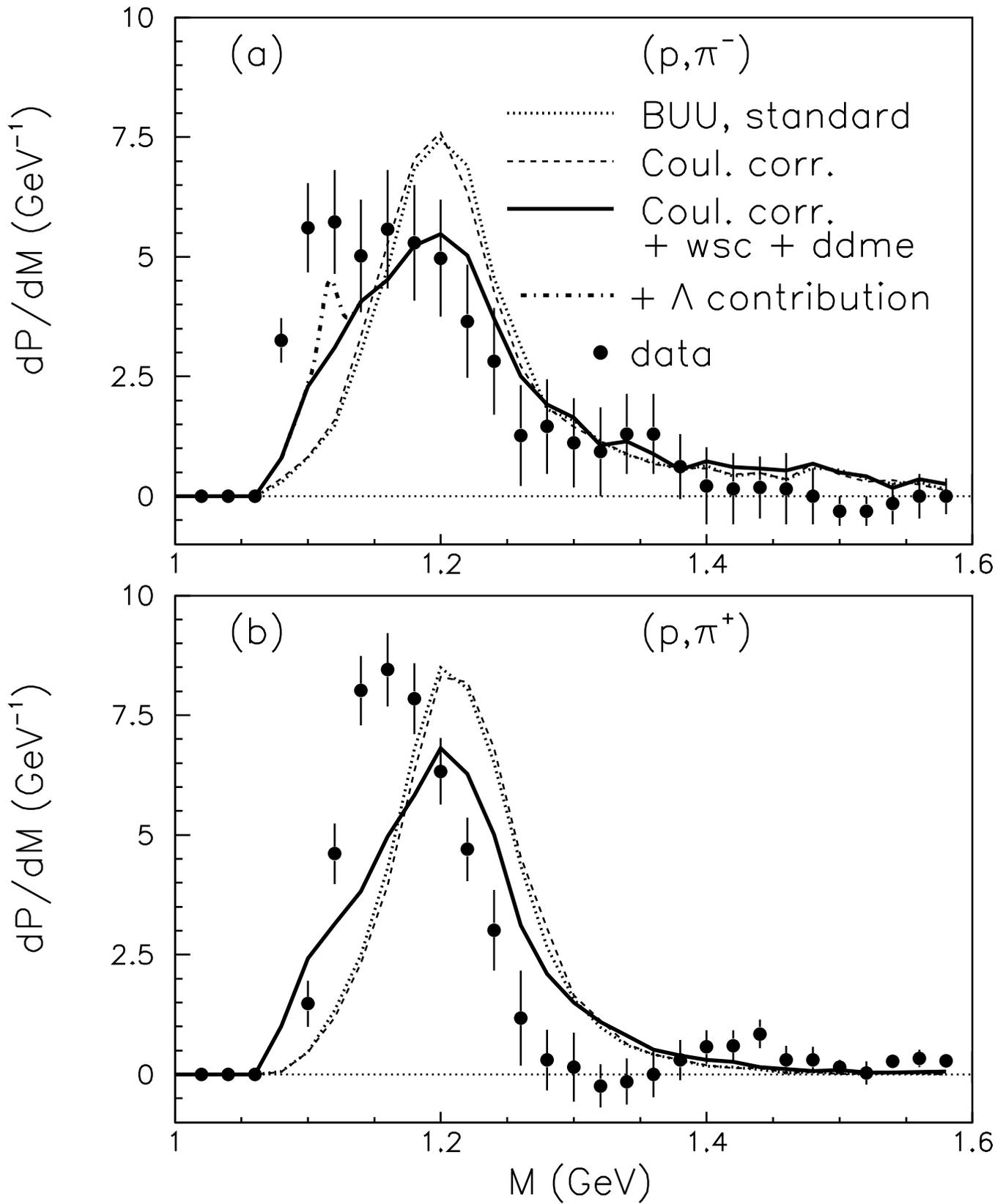
\item[Fig. 11] The same as Fig. 9, but for central collisions of  
Ni+Ni at 1.93 AGeV. The dot-dashed line shows the $\Lambda$ decay
contribution to the spectrum of $(p,\pi^-)$ pairs added to the 
$\Delta^0$ decay spectrum calculated with selfconsistent width of 
the $\Delta$-resonance and density-dependent matrix element.

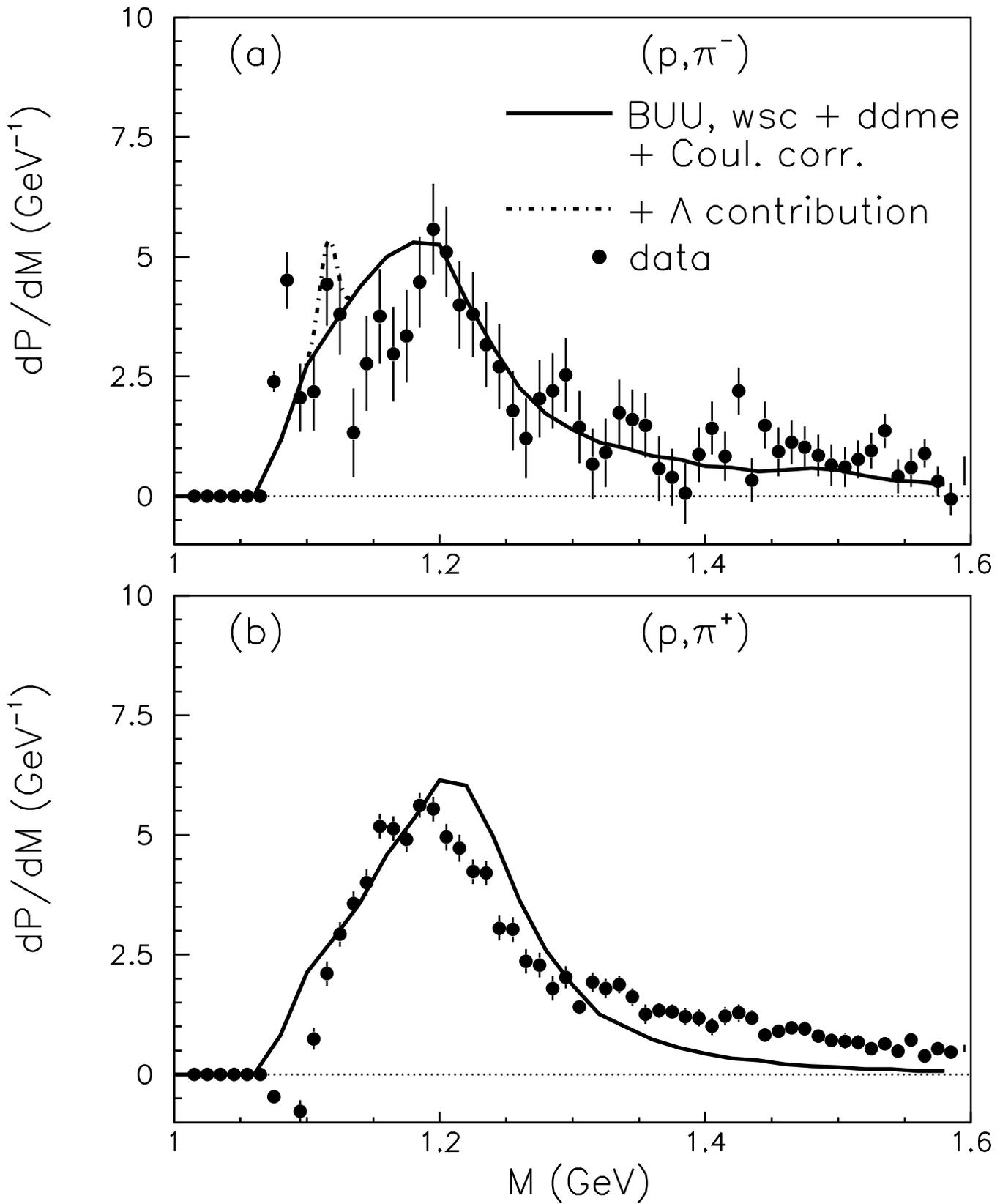
\item[Fig. 12]  The same as Fig. 9, but for Ni+Cu collisions
at 1.97 AGeV. The BUU calculation is performed with
selfconsistent $\Delta$-width and density-dependent matrix element
for the impact parameter region b=$0\div10$ fm. The final state Coulomb
interaction is taken into account.
The $\Lambda$ decay contribution to the spectrum of $(p,\pi^-)$ pairs
added to the $\Delta^0$ decay spectrum is shown by the dot-dashed line.
The experimental data are from Ref. \cite{Hjort97}.

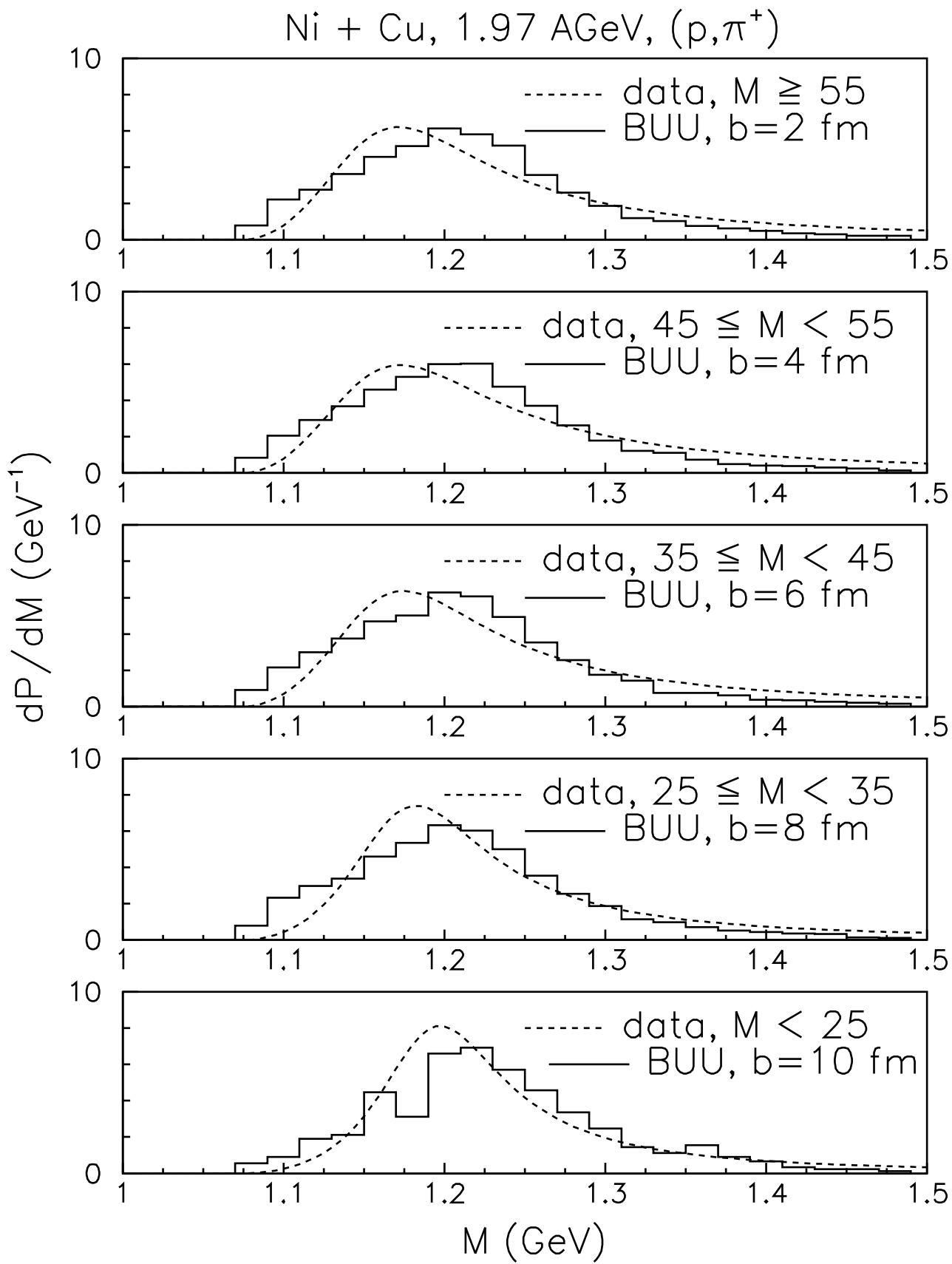
\item[Fig. 13] Invariant mass spectra of $(p,\pi^+)$ pairs for
Ni+Cu collisions at 1.97 AGeV at different centrality.
The BUU calculations (performed with selfconsistent $\Delta$-width 
and density-dependent matrix element)
are presented by histograms for b=2,4,6,8 and 10 fm from the uppermost to 
lowermost panel. The Breit-Wigner fits to the data from Ref. \cite{Hjort97}
are shown by dashed lines for the multiplicities $M\geq55$, $45 \leq M < 55$,
$35 \leq M < 45$, $25 \leq M < 35$ and $M < 25$.
The spectra are normalized to unity.

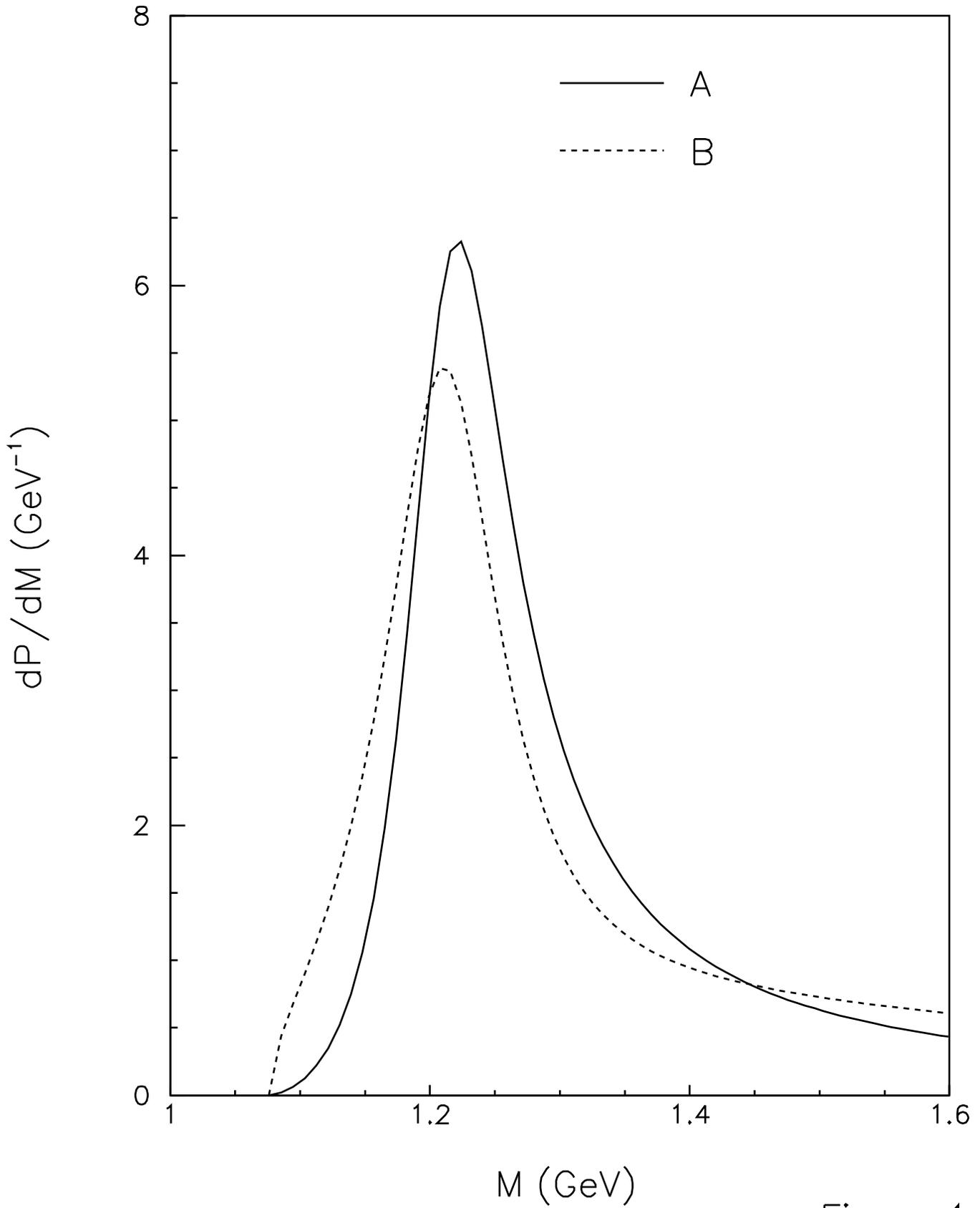
\item[Fig. 14] The spectral function ${\cal A}$ of the free $\Delta$
(solid line) and the weight function ${\cal B}/2\pi$ (dashed line) 
versus invariant mass. 

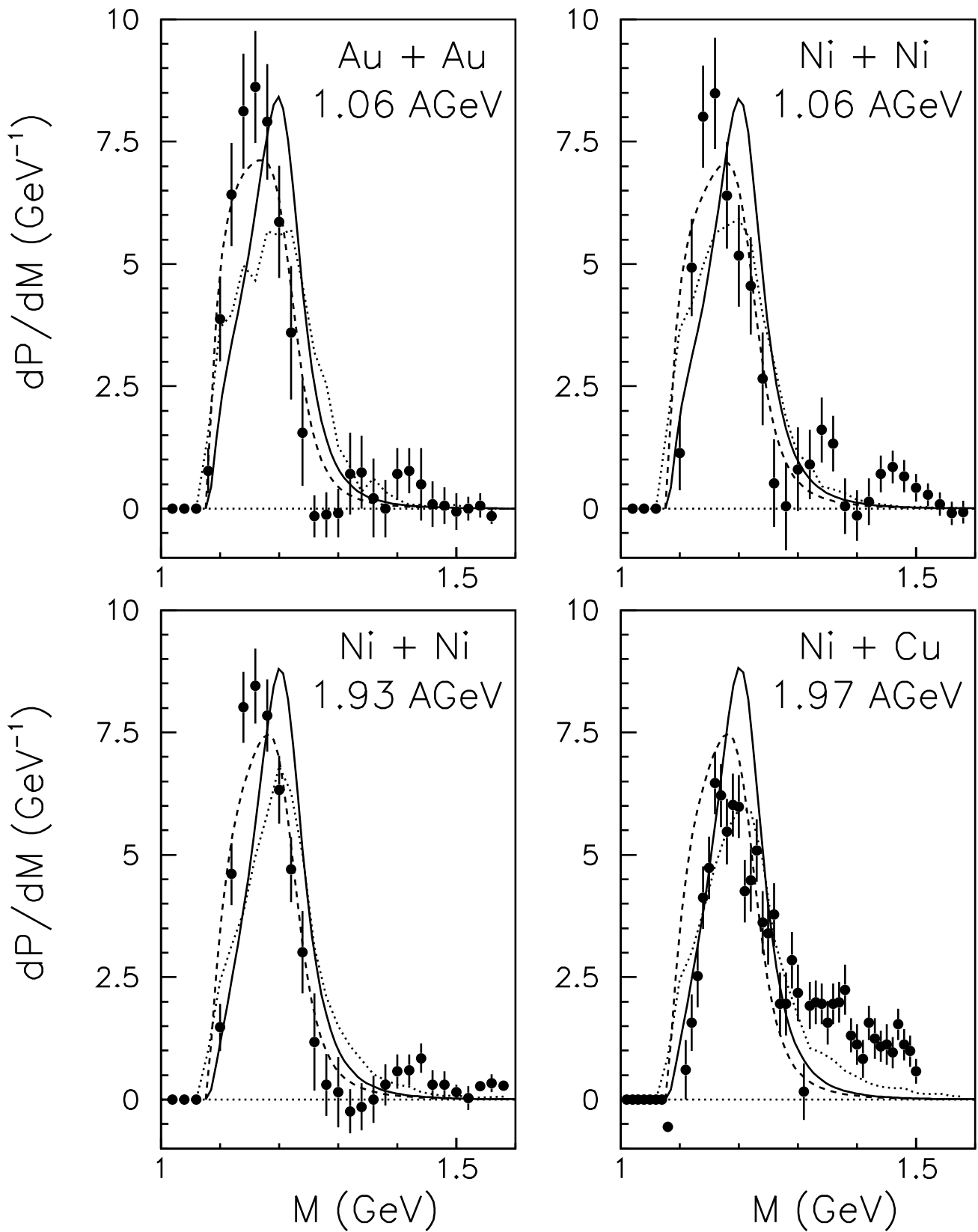
\item[Fig. 15] Thermal invariant mass distributions  using functions 
${\cal A}$ (solid lines) and ${\cal B}$ (dashed lines) weighted according 
to Eq. (\ref{pairaver}) in comparison to the data on $(p,\pi^+)$ pairs
from Ref. \cite{Eskef98} (Au+Au and Ni+Ni central collisions) and from
Ref. \cite{Hjort97} (Ni+Cu collisions with multiplicity 
$M \geq 55$). The dotted lines show the BUU (wsc+ddme) results for central 
collisions. All curves are normalized to unity.

\end{description}

\end{document}